\documentclass[3p,sort&compress]{elsarticle}

\usepackage{lineno,hyperref}
\hypersetup{colorlinks = true, allcolors = cyan}

\usepackage{amsmath}
\usepackage{makecell}

\def\mywidth{0.9\textwidth}

\journal{International Journal of Heat and Mass Transfer}

\newcommand{\Nu}{{\mathrm{Nu}}}
\newcommand{\Kn}{{\mathrm{Kn}}}
\newcommand{\Gr}{{\mathrm{Gr}}}
\newcommand{\Ra}{{\mathrm{Ra}}}
\newcommand{\Pra}{{\mathrm{Pr}}}
\newcommand{\Bi}{{\mathrm{Bi}}}

\begin{document}

\begin{frontmatter}

\title{A free convection heat transfer correlation for very thin horizontal wires in rarefied atmospheres}


\author[mymainaddress]{Lilian Peinado}

\author[mysecondaryaddress]{Victor Muntean\corref{mycorrespondingauthor}}
\cortext[mycorrespondingauthor]{Corresponding author}
\ead{victor.muntean@upm.es}

\author[mymainaddress]{Isabel Pérez-Grande}

\address[mymainaddress]{IDR/UPM, Universidad Politécnica de Madrid, Plaza Cardenal Cisneros 3, 28040 Madrid, Spain}
\address[mysecondaryaddress]{ETSIAE, Universidad Politécnica de Madrid, Plaza Cardenal Cisneros 3, 28040 Madrid, Spain}

\begin{abstract}
Mars and upper Earth atmosphere missions are booming in recent years increasing the interest in heat transfer in rarefied gases. At these low pressures, the heat transfer problems involving small geometries whose characteristic length is of the order of the mean free path of the gas molecules, raises new difficulties. This being especially severe in the case of free convection, where the velocity of the gas is closely linked to the heat transfer problem. Experiments with horizontal wires with diameters of {$12.7\ \mu\mathrm{m}$} and {$25\ \mu\mathrm{m}$} were conducted in a tailored vacuum chamber with pressures from {$0.03$~milibar} to ambient pressure, reaching temperature differences of up to {$80$~K} between the wire and the air. Experimental data is comprised of Rayleigh numbers from $10^{-13}$ to $10^{-5}$ that cover a wide range from free molecular flow to continuum regime. Particularly for the transition regime, with Knudsen numbers ranging between 0.1 and 10, the experimental data shows that the free convection heat transfer cannot be represented solely as a function of the  Rayleigh number, but a dependence on Knudsen number must be introduced into the correlation as well.  The presented results for the Nusselt number below  Rayleigh numbers of $10^{-8}$ exhibit a clear departure from the available correlations in the literature. A new empirical correlation for free convection from horizontal thin wires from the transition to continuum regime is presented. The present work sheds light on current thermal engineering problems which involves very thin wires in their systems which interact with a surrounding environment with a significant level of rarefaction. Systems such as temperature sensors based on thermocouple wires for Mars atmosphere whose measurements are affected by a low free convection heat transfer, could benefit from this study gaining a better performance in modelling with a more accurate estimation of free convection heat transfer.

\end{abstract}

\begin{keyword}
Heat transfer \sep Free convection \sep Rarefied atmospheres \sep Transition regime \sep Thin wires
\end{keyword}

\end{frontmatter}

\nolinenumbers

\section{Introduction}

The modelling of heat exchange between a solid system and a rarefied atmosphere is nowadays of great interest, due mainly to the increasing number of applications. Rarefied gases are found in other planets' atmospheres (robotic missions to Mars have to deal with this environment~\cite{gomez2012}), or in the upper layers of the Earth’s atmosphere where stratospheric balloons operate~\cite{perez2011}, as well as in ground experiments performed in controlled atmospheres~\cite{gao2018}. This matter is of special interest in small geometries where the flow may no longer be in a continuum regime but in a slip flow or transition regime towards the molecular regime~\cite{tsien1946,springer1971,davis1972}. This is for instance the case of the very thin wires (with a diameter of about 75 micrometers) used to build thermocouples to be part of the sensoring systems onboard Mars rovers~\cite{perez2017}.

In order to correctly model the heat exchange in these particular situations, the heat transfer coefficients have to be determined from the known correlations, either based on experimental measurements or CFD results~\cite{kyte1953, collis1954, mahony1957, fujii1979, fujii1982}. However, when reviewing the literature, one can see that there is no consensus in the estimation of convection through a rarefied gas. Furthermore, for very low pressure environments (transition regime) there is no correlation for thin wires published so far.

For this reason, the aim of this work is to determine the particularities of the heat transfer mechanisms for very thin wires (with diameters in the order of micrometers) in rarefied atmospheres. In these cases, the wires interact with the surrounding atmosphere in a slip flow or transition regime towards the molecular regime. Such regime or rarefaction degree can be determined by means of the Knudsen number, $\Kn$. Although this fact affects both forced convection~\cite{xie2017,xie2018} and free convection, the present study is focused on the experimental determination of the free convection coefficient of very thin horizontal wires in a wide range of Rayleigh numbers, $\Ra$, from $10^{-13}$ (molecular regime) to $10^{-5}$ (continuum flow). The extension of the experiments to the already known continuum flow has as its main aim the validation of the experimental setup and measuring system by comparison of the data here obtained with the results previously obtained by other researchers.

By studying the related published bibliography, one can find a number of experimental and analytical correlations to obtain the Nusselt number, $\Nu$, as a function of the $\Ra$ for horizontal cylinders or wires~\cite{morgan1975, churchill1975, boetcher2014}. They cover diverse fluid and gases as well as different aspect ratios $L/D$ of the wires. However, the first finding in this review is the scattering of data, in such a way that there is not a unique correlation that fits all experimental data, even in the same range of $\Ra$~\cite{kyte1953,fujii1982}. Moreover, none of them show experimental values for $\Ra < 10^{-10}$ for horizontal wires. Consequently, the present work covers the gap in the transition regime. Based on the measurements described in the following sections, an experimental correlation has been obtained in the entire range $10>\Kn>0.1$. The main difference in the new proposed correlation with respect to previous works relies on the estimation of $\Nu$ as a function, not only of the $\Ra$, but also of the $\Kn$, achieving an error lower than {$10\%$} with experimental data.

This new correlation can be applied in the different scenarios described above: convection heat exchange of horizontal fine wires in rarefied planetary atmospheres, hot wire anemometry in rarefied gases or stratospheric conditions, etc. Furthermore, this study can support thermal design in micro and nano-technology based on fine wires which operate at high Knudsen numbers~\cite{seo2019}.

The following sections describe the experimental setup, the measurements, and the correlation obtained, including the validation of the experimental data with results of previous research works.

\section{Experimental setup}

The experiments were performed in a vertical cylindrical custom built vacuum chamber, made of acrylic glass. The external diameter of the cylinder is {$300$~mm}, its height {$400$~mm} and the thickness of the wall {$4$~mm}. It is closed at both ends by means of two flat acrylic plates with a wall thickness of {$15$~mm}. The top cover is welded using trichloromethane, and additional sealing is achieved by adding {Araldite 2004} epoxy adhesive at the joints. The bottom plate is removable, and the sealing is achieved using an 8 mm rubber O-ring. This plate has four mechanized openings  compatible with standard {DN 25 ISO-KF} flanges.

One of the openings  is used to connect the vacuum pump ({Leybold~Scrollvac~18}). As the pump does not have the capability of pressure control, the desired vacuum level is achieved by operating the vacuum chamber with a controlled leakage in the pumping line by means of a needle valve.  Once the desired level of vacuum is achieved, the pumping line is closed and the pressure inside the chamber remains steady. The pressure is measured using a pressure sensor ({Thermovac~TTR911}) connected to the second opening.

\begin{figure}[ht]
	\includegraphics[width=\mywidth]{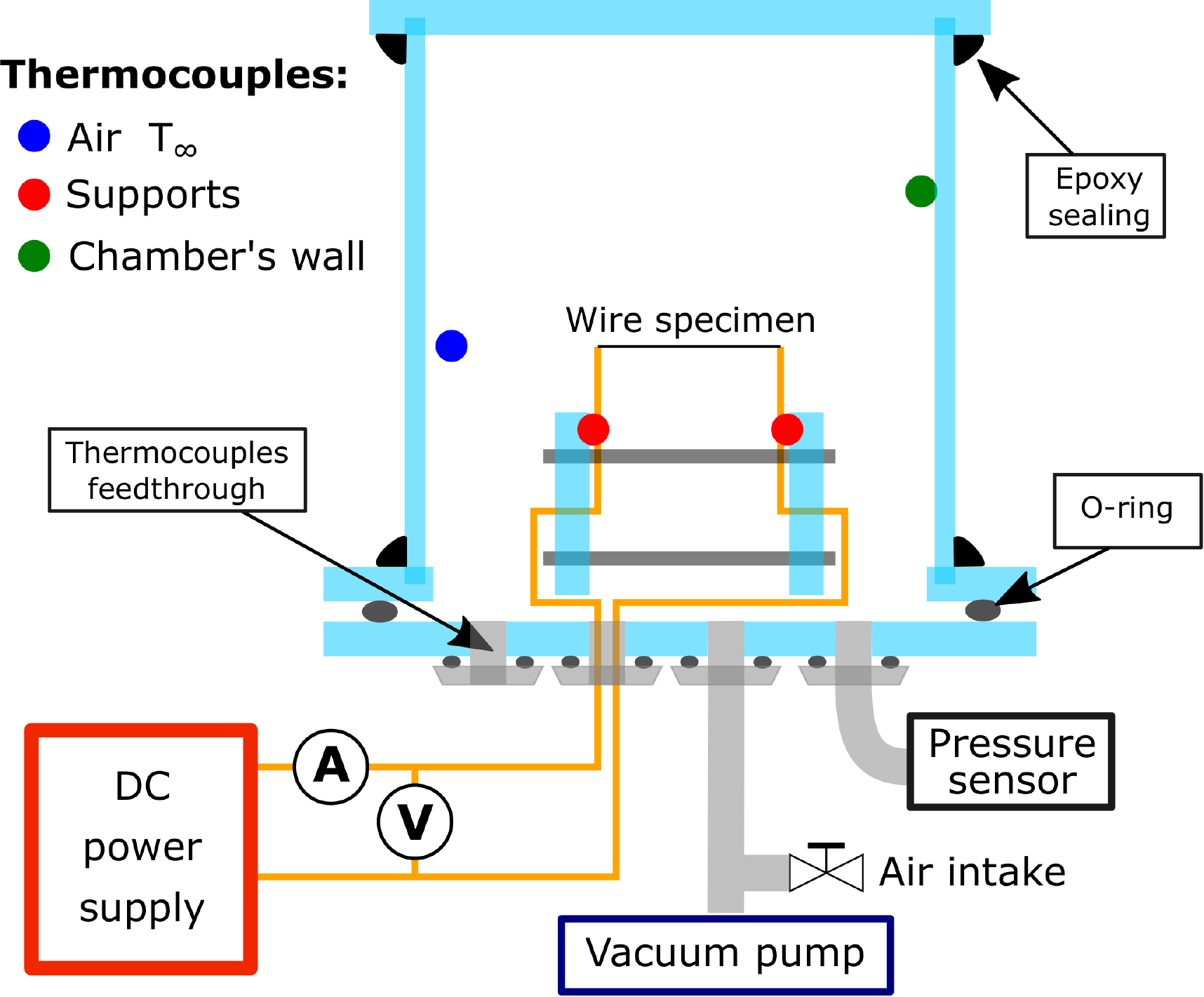}
	\centering
	\caption{Scheme of the vacuum chamber and experimental setup.}
	\label{fig:setup}
\end{figure}

The other two holes are used as electrical feedthroughs. The first one is used to connect the thermocouples placed inside the chamber with the acquisition system on the outside. Four {type-T} thermocouples were used to measure the temperatures of the air far from the wire in its horizontal line, of the supports of the heated wire and of the chamber's wall. The second feedthrough is used to connect the DC power supply to the electrodes for supplying electrical current to the heated wire. A simplified scheme of the described experimental setup is presented in the Figure \ref{fig:setup}.

The experiments were performed using alumel wires. Alumel is an alloy with an approximate composition $95\%$ nickel, $2\%$ aluminium, $2\%$ manganese, and $1\%$ silicon. This kind of wires are used as the negative conductor of \mbox{type-K} thermocouples. The diameters of the wires used in our experiment were {$12.7\ \mu\mathrm{m}$} and {$25\ \mu\mathrm{m}$}, with lengths of {$61$~mm} and {$64$~mm}, respectively. In the experimental setup the wires are soldered to two copper electrodes horizontally. These electrodes have a diameter of {$2$~mm} and are assembled in a structure support built for the test and placed inside the chamber. The connection to the power source is made using an electrical feedthrough as mentioned above.

The electrical connections outside the chamber allow the supply of a constant current from a variable DC current supply to the test assembly, as well as the measuring of the voltage at the entrance of the chamber, and the electrical current through the wire's section by means of two {Keithley~2000} multimeters. The wiring of the electrical setup was characterized in order to obtain its electrical resistance.

The values of voltage and electrical current, as well as pressure and temperature were monitored and recorded in the computer during the experiments using a data acquisition platform (National Instruments CompactDAQ).

\section{Problem statement}
\label{sec:prob}
\subsection{Wire's temperature}

Prior to stating the heat transfer problem, the procedure to determine the wire’s temperature is described, as it will be necessary to calculate the heat exchange coefficient between the wire and the surrounding air. A wire’s temperature can be obtained from the known dependence between the electrical resistance of a wire and its temperature,
\begin{equation}
R_w = R_0[1+\alpha(T_w-T_{w0})],
\label{eq:Tw}
\end{equation}
where $R_w$ and $T_w$ are the electrical resistance and the temperature of the wire, the subscript $0$ denotes a known reference state, and $\alpha$ is the temperature coefficient. This linear approximation is suitable for the range of the wire's temperatures used in our experiments which was between $293$ and {$380$~K}~\cite{kasap2017}. The value of the temperature coefficient of this material in this range was obtained experimentally. The result was {$\alpha=0.0023\ \mathrm{K}^{-1}$}, which is in line with the value {$0.00239\ \mathrm{K}^{-1}$} provided by the manufacturer, Omega\textsuperscript{TM}.

For the heat transfer experiments the wire is heated by Joule effect. A set of $95$~tests were performed using the wire with the diameter of {$25$~$\mu\mathrm{m}$}. The employed pressures were between $0.03$~milibar to ambient pressure, and $7$~different current levels were used ranging from $2$ up to $38$~milliamperes  Each  experiment was performed at a constant pressure, and the selected current level was kept constant until the variations of the temperature difference between the wire and the far away air were below {$0.1$~K} for at least 15~minutes. Once fulfilled this criterion the steady state is achieved, and the voltage  drop between the electrode leads and the electrical current passing through the wire are  measured and used to compute the actual electrical resistance of the wire. Ultimately, the wire's mean temperature is obtained by solving the equation~(\ref{eq:Tw}).

\subsection{Theoretical considerations}
In order to calculate accurately the heat transfer between a body and the surrounding atmosphere, it is necessary to solve the Navier-Stokes equations in the medium, which is very expensive in terms of computing and analytic solutions are feasible only for very simple geometries. In engineering applications this process is avoided and empirical correlations for the most common configurations are used instead. These correlations are expressed as a function of the non-dimensional numbers involved in the problem. It is well known that for a given geometry the natural convection heat rate can be obtained by knowing the value of the Nusselt number as a function of Rayleigh number, which can be expressed as the product of Grashof and Prandtl numbers~\cite{incropera}, whose product is the Rayleigh number, $\Ra = \Gr\Pra$. Whilst the value of $\Pra$ is almost constant and near $0.7$ for air in the range of interest, the behaviour of Nusselt can be described properly either by using the Rayleigh or Grashof number.

For the case of a slender enough horizontal cylinder of length~$L$, and diameter~$D$, (${L/D\gg1}$), the representative dimension of the problem is its diameter, thus it is the magnitude to be used in the expressions of the non-dimensional numbers. Insofar the Nusselt number is 
\begin{equation}
\Nu_D = \frac{hD}{k},
\label{eq:nu}
\end{equation}
where $h$ is the mean convection heat transfer coefficient. It does not account for azimuthal angle with respect to the vertical plane  or for longitudinal variations. The convection coefficient is calculated using the heat rate loss by convection from the wire for a given temperature difference,
\begin{equation}
h = \frac{q_{conv}}{\pi D L \Delta T}.
\label{eq:coeff}
\end{equation}
The value of $h$ varies significantly with the diameter of the wire, thus for very thin wires it can reach values up to $800\ \mathrm{W/m^2K}$.

The Grashof number is defined as
\begin{equation}
\Gr = \frac{g\beta \Delta T D^3}{\nu^2},
\label{eq:gr}
\end{equation}
where $\beta$  is the coefficient of thermal expansion, which for the ideal gas hypothesis is $\beta = T^{-1}$~\cite{incropera}, $\nu$ is the kinematic viscosity, and $k$ the thermal conductivity of the air. These three magnitudes are temperature dependent  and are evaluated at the average temperature of the air and the wire, or film temperature, defined as  ${T_f = 0.5(T_w+T_{\infty})}$.

It is important to note that for the characteristics of the wires used, and with temperature differences of about ${10\ \mathrm{K}}$, the Grashof number is estimated to be in the order of $10^{-5}$. Reducing the pressure in the vacuum chamber to obtain rarefied atmospheres decreases this value even more, as the Grashof number is directly proportional to the square of the density. Whereas for Grashof values significantly higher  than this case there are lot of correlations available in the literature, for very thin wires there are very few. The most representative ones proposed by \citet{collis1959}, \citet{kyte1953}  and \citet{fujii1982} for the range of Grashof numbers of interest in this paper are presented in Table \ref{tab:corr}.

\begin{table}
	\centering
	\caption{Correlations}
	
	\begin{tabular}{ c c l}
		\hline
		Correlation & Validity range & Reference \\ \hline 
		 & & \\
		$\dfrac{2}{\Nu_D} = 1.627 - 0.86 \log_{10} \Ra$ & $10^{-10} < \Ra < 10^{-2} $ & \citet{collis1959} \\ 
		 & & \\
		$\dfrac{2}{\Nu_D} = \ln \left( 1.0 + \dfrac{7.09}{\Ra^{0.37}}\right)$ & $10^{-7} < \Ra < 10^{1.5} $ & \citet{kyte1953} \\ 
		 & & \\  
		\makecell{ $\dfrac{2}{\Nu_D} = \ln \left( 1.0 + \dfrac{3.3}{C(\Pra) \Ra ^n} \right)$  \\ $C(\Pra) = 0.671/ \left\{1+(0.492/\Pra)^{9/16}\right\}^{4/9} $ \\ $n = 0.25+ 1.0/(10+5\Ra^{0.175}) $ } & $10^{-8} < \Ra < 10^{6} $ & \citet{fujii1982} \\  
		 & & \\ \hline
		
	\end{tabular}
\label{tab:corr}
\end{table}

\subsection{Heat transfer estimation}
When the steady state conditions are attained, the electrical power supplied to the wire, $q_\mathrm{J}$, is lost by convection, $q_\mathrm{conv}$, conduction, $q_\mathrm{cond}$, and radiation, $q_\mathrm{rad}$. Therefore, the convective heat flow can be obtained from
\begin{equation}
q_\mathrm{conv} = q_\mathrm{J} - q_\mathrm{cond} - q_\mathrm{rad}.
\label{eq:balance}
\end{equation}

The power supplied to the wire, $q_\mathrm{J}$, can be modelled with a simple electrical circuit constituted by two series mounted electrical resistors: the wire resistance, $R_w$, and the auxiliary wiring  electrical resistance, $R_c$. This last resistance is  constant for all the experiments, thus $q_\mathrm{J}$ can be calculated from the voltage,$V$, and current, $I$, measured at the entrance of the vacuum chamber. It can be written as
\begin{equation}
q_\mathrm{J} = VI-R_cI^2.
\label{eq:joule}
\end{equation}

In several experimental works~\cite{collis1954,kyte1953,milano2010}, the heat losses by conduction to the supports of the wire are assumed to be negligible. This is acceptable for pressures close to the atmospheric pressure. However, the relative importance of the conductive losses increases as the pressure decreases. In order to reduce the potential errors caused by neglecting these losses, the solution of the {one-dimensional} heat-conduction model of the wire, assuming an average heat transfer coefficient, is used~\cite{incropera}. The conduction losses in this model are
\begin{equation}
q_{cond} = \frac{q_J\tanh(\Lambda \sqrt{\Bi})}{\Lambda\sqrt{\Bi}},
\label{eq:cond}
\end{equation}
where $\Bi$ is the Biot number and $\Lambda$ is the aspect ratio of the wire, $L/D$. The Biot number can be expressed as the product between the Nusselt number and the air and wire's conductivities ratio, $\Bi = \Nu_D(k/k_{w})$. The calculations show that the losses are lower than the $3\%$ of the applied electrical power for pressures above {$10$~mbar}, but rapidly increase up to $8\%$ when the pressure decreases further. This value is similar to the one calculated by \citet{milano2010}, whose 3D model showed losses up to $6\%$ of the total dissipated heat.

The thermal radiation losses  are estimated using 
\begin{equation}
q_{rad} = \varepsilon  \pi DL \sigma(T_w^4-T_{c}^4),
\label{eq:rad}
\end{equation}
where $\sigma$ is the Stefan-Boltzmann constant and $\varepsilon$ is the wire's infrared emissivity, which is assumed to be $0.08$~\cite{sasaki1994}. The maximum value of the thermal radiation losses was lower than the {$2\%$} of the total  electrical power dissipated by the wire.

Once the convective heat rate has been calculated from equation~(\ref{eq:balance}),the average convection heat transfer coefficient $h$ is obtained from equation~(\ref{eq:coeff}).

\section{Experimental procedure}
More than $200$ tests were performed to check the repeatability of the experiment. A summary of the values used is shown in Table \ref{tab:test}.

\begin{table}
	\centering
	\caption{Summary of values of experimental parameters}

	\begin{tabular}{ll}
		\hline
		Variable & Value  \\ \hline 
		Wire diameter & {$12.7\ \mu \mathrm{m}$}, {$25\ \mu \mathrm{m}$} \\
		Wire length & {$61\ \mathrm{mm}$}, {$64\ \mathrm{mm}$} \\
		Chamber pressure & 0.03 to 942 mbar \\
		Supplied power & 0.4 to 145 mW \\
		Wire's temperature & 296 to 371 K \\ \hline
				
	\end{tabular}
	\label{tab:test}
\end{table}

Experiments using the {$25$~$\mu \mathrm{m}$} diameter wire were performed in the first place. Due to the lack of data in the range of Grashof numbers  of present interest, it would not be possible to compare our results with other author's, except for some experimental points located in~\cite{milano2010}. For this reason, more experiments were performed using the {$12.7\ \mu \mathrm{m}$} wire to validate the  correlation obtained from the results of the  $25\ \mu \mathrm{m}$ wire. Every experimental case was run at constant preselected pressures, ranging from $0.03$ to $942$ mbar. The wire was supplied with electrical power from the DC power source at constant current mode until the steady state was achieved. 

For every experimental run the temperature of the wire is calculated using equation~(\ref{eq:Tw}). Then film temperature, $T_f$, is obtained, and the density is calculated using the ideal gas equation. Next, the viscosity, $\mu$, as well as the gas thermal conductivity, $k$, are obtained from tabulated data~\cite{tablas}. Afterwards, the Rayleigh number of each run is calculated using the Grashof number, equation~(\ref{eq:gr}), and the Prandtl number defined as $\mathrm{Pr} = \mu c_p/k$.

Once all the parameters are calculated, the heat rate supplied to the wire is calculated using equation~(\ref{eq:joule}). Radiation losses are estimated using equation~(\ref{eq:rad}), where all the variables are known. For the estimation of conduction losses to the wire's supports, the value of $\Nu_D$ is needed. At this point, this value is not known so an iterative process is used starting with a typical value of Nusselt number for thin wires, $\Nu_D = 0.3$. With this estimation and the balance equation~(\ref{eq:balance}) the convective heat rate is calculated, which gives the value of the Nusselt number using equations~(\ref{eq:coeff}) and (\ref{eq:nu}). Since this is not the actual value, it is used as a new input for the estimation of the conduction losses. The process is considered to converge when the difference between the two Nusselt numbers is less than $0.1\%$. It has converged in most cases in less than five iterations.

For the estimation of the uncertainty in the values of the Nusselt number, the absolute error introduced by measuring the air's temperature using a thermocouple is estimated at $1~\mathrm{K}$. The wire's length measurement introduces an error of $1~\mathrm{mm}$. Whereas, for the values of $I$ and $V$ conservative values of the relative errors of $1\%$ and  $0.1\%$ are used respectively. The relative error is $0.2\%$ for $R_c$ determination. All the errors propagate to the total uncertainty for the Nusselt number, which is calculated to be below $3\%$ following the methodology described by \citet{beckwith2009}.

\section{Results and discussion}
\begin{figure}[h]
	\includegraphics[width=\mywidth]{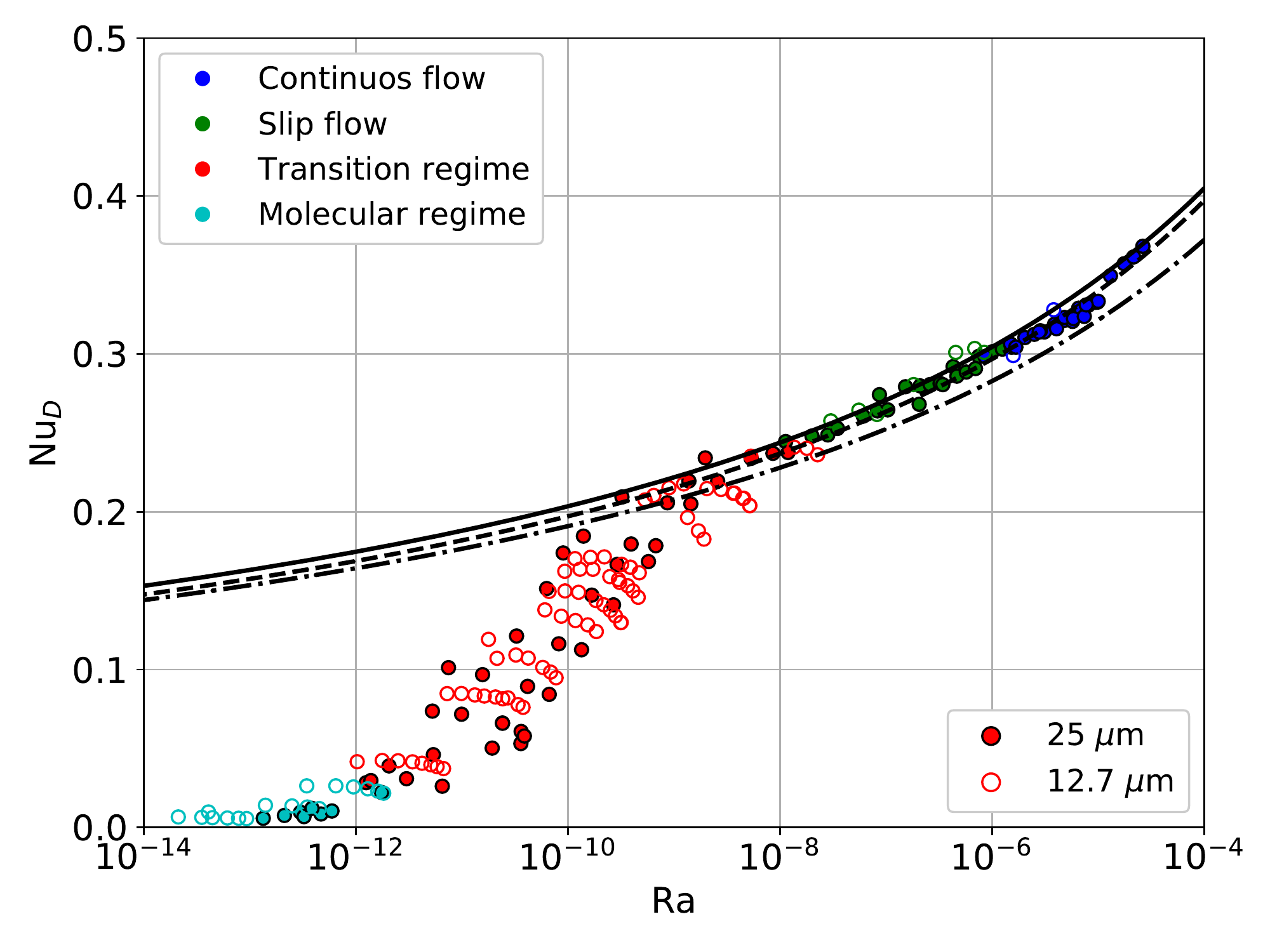}
	\centering
	\caption{Experimental results of the Nusselt number as a function of Rayleigh number. The results are compared with available correlations: \citet{fujii1982}(\textit{solid}), \citet{collis1959}(\textit{dashed}) and \citet{kyte1953}(\textit{dash-dotted}).}
	\label{fig:nu_exp}
\end{figure}

The results of $\Nu_D$ and $\Ra$ for horizontal wires of diameters {$12.7$~$\mu \mathrm{m}$} and {$25$~$\mu \mathrm{m}$} are presented in the graph shown in Figure \ref{fig:nu_exp}. In the same graph the  correlations  available presented in Table \ref{tab:corr} are also plotted. The experimental results for Rayleigh numbers higher than $10^{-8}$ are in good agreement with the correlations, even if there are differences in the estimations of  $\Nu_D$ between authors up to $10\%$.

\begin{figure}[h]
	\includegraphics[width=\mywidth]{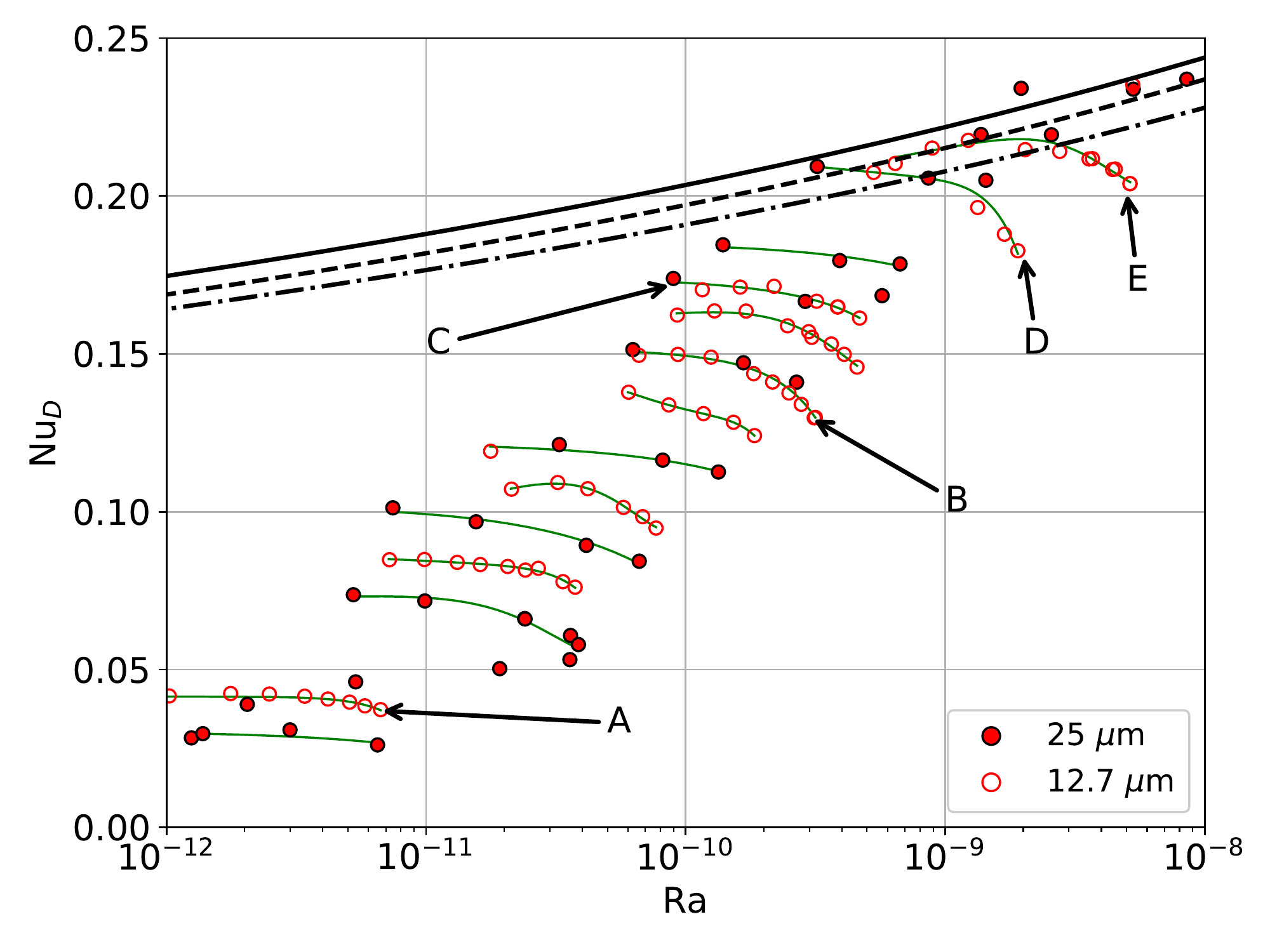}
	\centering
	\caption{Experimental results of the Nusselt number as a function of Rayleigh number in the range {$10^{-12}<\Ra<10^{-8}$}. \textit{Green} lines approximately represent constant values of $pD$ for every cluster of experimental points. Clusters \textit{A, B, C, D} and \textit{E} are formed by experimental results for both wire's diameter, {$12.7$~$\mu \mathrm{m}$}  and {$25$~$\mu \mathrm{m}$}. The available correlations are plotted: \citet{fujii1982}(\textit{solid}), \citet{collis1959}(\textit{dashed}) and \citet{kyte1953}(\textit{dash-dotted}).}
	\label{fig:nu_exp_pd}
\end{figure}

Out of the application range of the mentioned correlations, an abrupt drop of the Nusselt number is noticed at $\Ra$ below $10^{-8}$. The same behaviour in this region, where the effects of rarefied atmosphere are noticed, was pointed out by \citet{kyte1953} in their experiments with vertical wires. It can be observed that for these points the Rayleigh number is no longer enough to describe the behaviour because these experimental points cannot collapse over a continuum curve  as the ones in the range of Rayleigh higher than $10^{-8}$. In the range of $\Ra$ between $10^{-12}$ and $10^{-8}$, each cluster of experimental points obtained at the same pressure level seem to form a curve. The curves corresponding to a constant pressure do not overlap if different wire diameters are used. Although, as it is shown Figure \ref{fig:nu_exp_pd} where the mentioned range of $\Ra$ from Figure \ref{fig:nu_exp} is zoomed in, some of the clusters are formed by experimental results of both wire's diameter, {$12.7$~$\mu \mathrm{m}$}  and {$25$~$\mu \mathrm{m}$}. An in depth analysis of this behaviour, exposed that the mentioned clusters of points correspond to constant values of the product between pressure and wire's diameter. This conclusion suggests that in this region, the pressure and wire's diameter are needed independently from the Rayleigh number, in order to describe the heat transfer problem. As explained later, the product of these variables can be replaced by a non-dimensional parameter, the Knudsen number.

For Rayleigh numbers below $10^{-12}$ the experimental data seem to approach asymptotically to $\Nu_D = 0$ (Figure \ref{fig:nu_exp}). Values of Nusselt number as small as $0.005$ are reached. However, the heat transfer for this range of Rayleigh is dominated by free molecular conduction~\cite{springer1971} and its analysis is beyond the scope of this experimental work.

\begin{figure}[h]
	\centering
	\includegraphics[width=\mywidth]{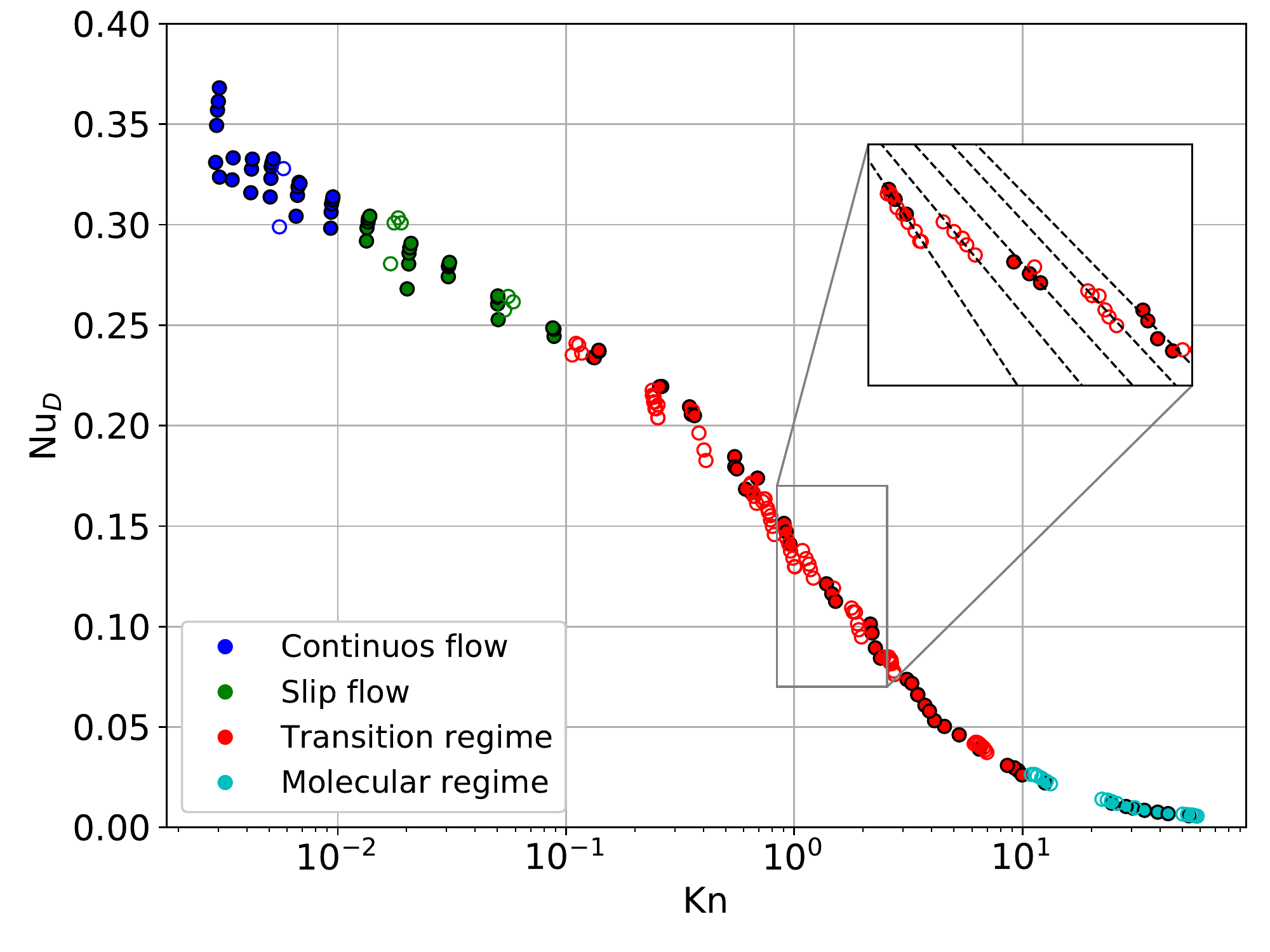}
	\caption{Experimental results of the Nusselt number as a function of Knudsen number. The detailed view shows clusters of experimental points aligned with non-parallel straight lines, indicating a dependency of $\Nu_D$ with another parameter different from $\Kn$ in the transition regime. These straight lines approximately represent constant values of $pD$ for every cluster of experimental points.}
	\label{fig:kn_nu}
\end{figure}

As the experiments were carried out in rarefied atmospheres, the Knudsen number can be introduced as a new variable that has an influence on the Nusselt number. The Knudsen number is defined as the relation between the mean free path of molecules, $\lambda$, and the characteristic length scale of the problem, the wire's diameter in this case, $\Kn = \lambda/ D$. The mean free path of molecules is
\[\lambda = \dfrac{\mu\sqrt{2\pi R_g T_f}}{2p},\]
where {$R_g = 287$~J/(kgK)} is the specific gas constant of air, $\mu$ is its viscosity, and $T_f$ and $p$ are the experimental values of film temperature and pressure respectively~\cite{xie2017}. In Figure \ref{fig:kn_nu}  the behaviour of the Nusselt number with the Knudsen number is presented. In this graph the $4$ regimes described by \citet{springer1971} are well defined:
\begin{itemize}
	\item Continuum: $\Kn < 0.01,$
	\item Slip flow: $0.01 < \Kn < 0.1,$
	\item Transition: $0.1 < \Kn < 10,$
	\item Free molecule: $\Kn > 10,$
\end{itemize}
although these limits are merely approximate. The described regimes are also presented in Figure \ref{fig:nu_exp}, where the dependency of experimental points on $\Kn$ is not obvious.

The pattern of the data shown in the Figure \ref{fig:kn_nu} suggests that there are three trends that can be distinguished. A low Nusselt zone ($\Nu_D<0.05$) for $\Kn > 5$ corresponding to the free molecule regime, a transitional zone, and a high Nusselt number ($\Nu_D>0.25$) zone. For the highest values of Nusselt  the difference between continuum regime and slip flow regime is imperceptible. Indeed as it is shown in the graph in the Figure \ref{fig:nu_exp}, the experimental points in continuum and slip flow are in good agreement with the correlations presented in Table \ref{tab:corr}. So, even if physically there are two different regimes, there is no need to look up for a correlation for each zone. It is also patent that the slip flow regime could be extended up to $\Kn \approx 0.2$. In this range of $\Kn$ it is evident too that for a given value of $\Kn$ there is more than one value of $\Nu_D$ aligned vertically, which is an indication that here the dependence of the Nusselt number with the Knudsen number must be negligible. Thus, the Rayleigh number is enough to describe the heat transfer process in this range.

Finally the zone with a different behaviour from the other two identified above is the transitional zone, for Knudsen numbers between $0.2$ and $5$. Here, the experimental points seem to collapse on a single curve, which would indicate a one-to-one correspondence between $\Kn$ and $\Nu_D$. However, as it is shown in the detailed view in Figure \ref{fig:kn_nu}, this is inaccurate. Actually, different clusters of experimental points are aligned with non-parallel straight lines. Each straight line corresponds with a constant value of the product between pressures and wire's diameter. This behaviour clearly suggests that there is at least one parameter apart from the Knudsen number that describes the heat transfer, namely the Rayleigh number.

Taking into account that the boundaries for the different regimes are not clearly defined, we will use the same limits proposed by \citet{springer1971}, bearing in mind that for free convection heat transfer correlations the separation between continuum and slip flows is not important.

\section{Correlation}

The main goal of the presented work is to find a way to predict the heat losses from very fine wires in rarefied atmospheres, similar to the Mars atmosphere. This particular condition is outside the pure molecular regime of heat conduction and belongs to the range integrated by the slip flow and transition regime. Thus, the applicability range of the sought correlation will be in the range of Knudsen numbers below 10.

As it was mentioned in the previous section, there is no noticeable difference in the Nusselt number trend between the continuum and the slip flow regimes. Conversely, in the transition regime the trend is remarkably different and in addition, the inclusion of the Knudsen number is needed to characterize  the heat flow.

In order to specify which form of the expression that  correlates the Nusselt number with the other non-dimensional parameters involved in the problem would be, a hypothetical case without gravity is used. The wire is modelled as an infinite cylinder of diameter $D$. In this virtual situation all the heat power is lost by pure conduction from the heated wire to the surrounding gas. This simple problem is solvable only if a boundary condition is given such as a known temperature at a fixed distance from the wire. Let's suppose that the diameter of this boundary is $D_\mathrm{b}$ and its temperature is the far away temperature, the same as in the original convection problem, $T_{\infty}$. The solution is trivial and the heat loss rate per unit of length  from the wire would be
\begin{equation}
q' = 2\pi k \frac{T_w-T_{\infty}}{\ln (D_\mathrm{b}/D)}.
\end{equation}
Consequently, the following equation can be used to calculate the Nusselt number:
\begin{equation}
\frac{2}{\Nu_D} = \ln ({D_\mathrm{b}}/{D}).
\label{eq:nu_cond}
\end{equation}

In the original problem, a motion in the gas arises due to the presence of gravity and the density variations within the gas. Hence, the hypothetical cylindrical boundary with a given diameter and at a constant temperature cannot exist. The actual shape of the isotherm $T_{\infty}$ will be strongly related to the flow structure around the wire. In other words, this boundary will be a function of $\Ra$ for the continuum and  slip flow regimes and of $\Ra$ combined with $\Kn$ for the transition regime. Therefore, the sought correlations will have the form
\begin{equation}
\frac{2}{\Nu_{D,t}} = \Psi_t (\Ra , \Kn)
\end{equation}
for the transition regime, and the form
\begin{equation}
\frac{2}{\Nu_{D,c}} = \Psi_c (\Ra)
\end{equation}
for Knudsen numbers lower than $0.1$. Evidently, $\Psi_t$ and $\Psi_c$ are non-dimensional functions and, as seen in equation~(\ref{eq:nu_cond}), the natural logarithm is advisable to be used. 

Bearing in mind the discussion above, for the continuum regime the sought correlation will have a form similar to the one proposed by \citet{collis1959},
\begin{equation}
\frac{2}{\Nu_{D,c}} = A_c + B_c\ln(\Ra).
\label{eq:corr_cont}
\end{equation}
The constants $A_c$ and $B_c$ are calculated from experimental data by plotting the Nusselt values for $\Kn<0.1$ versus the $\ln(\Ra)$ and using the least mean square method for the fitting. The obtained values are $A_c = 2$ and $B_c = -0.34$. As can be observed in table \ref{tab:corr}, these coefficients are similar to the ones proposed by \citet{collis1959}.

As has been mentioned, for the characterization of the Nusselt number in the transition regime the introduction of the Knudsen number is advisable in the sought correlation. In this regime the mean free path of the molecules begins to be comparable to the wire diameter itself. We decided to use for the definition of non-dimensional parameters the modified diameter used by \citet{kyte1953}, $D_\mathrm{m} = D + 2\lambda$. It is trivial to relate this diameter with the Knudsen number using the expression $D_\mathrm{m} = D(1+2\Kn)$, whilst $\Kn = \lambda/D$. It is reasonable to use the modified diameter as a characteristic dimension of the problem, insomuch as in the region between the wire's wall and the distance $\lambda$ from the wall, where there are few collisions between gas molecules and the local thermodynamic equilibrium is not attained. Thus, it can be supposed that the temperature at a distance $\lambda$ from the wire's wall will be in close proximity to $T_w$. In this case, the modified Grashof number is defined using a wire with a diameter $D_\mathrm{m}$ whose wall is at a temperature $T_w$,
\begin{equation}
\Gr_\mathrm{m}  = \dfrac{g\beta (T_w-T_\infty)D_\mathrm{m}^3}{\nu^2} = \dfrac{g\beta (T_w-T_\infty)D^3(1+2\Kn)^3}{\nu^2},
\end{equation}
thus the modified Grashof number will be
\begin{equation}
\Gr_\mathrm{m} = \Gr(1+2\Kn)^3.
\end{equation}
Finally, the modified Grashof number, which would be the actual one in the described flow configuration in the vicinity of the wire, can be replaced by the product between the Grashof based in the actual diameter of the wire and a simple function of the Knudsen number.

Using the same process for the Nusselt number the modified value can be expressed as follows
\begin{equation}
\Nu_{D,\mathrm{m}} = \dfrac{hD_\mathrm{m}}{k} =\frac{hD}{k}(1+2\Kn) = \Nu_{D,t}(1+2\Kn).
\end{equation}
Therefore, the modified Nusselt number can be expressed as the product between the transitional Nusselt number and a function of the Knudsen.

\begin{figure}[h]
	\centering
	\includegraphics[width=\mywidth]{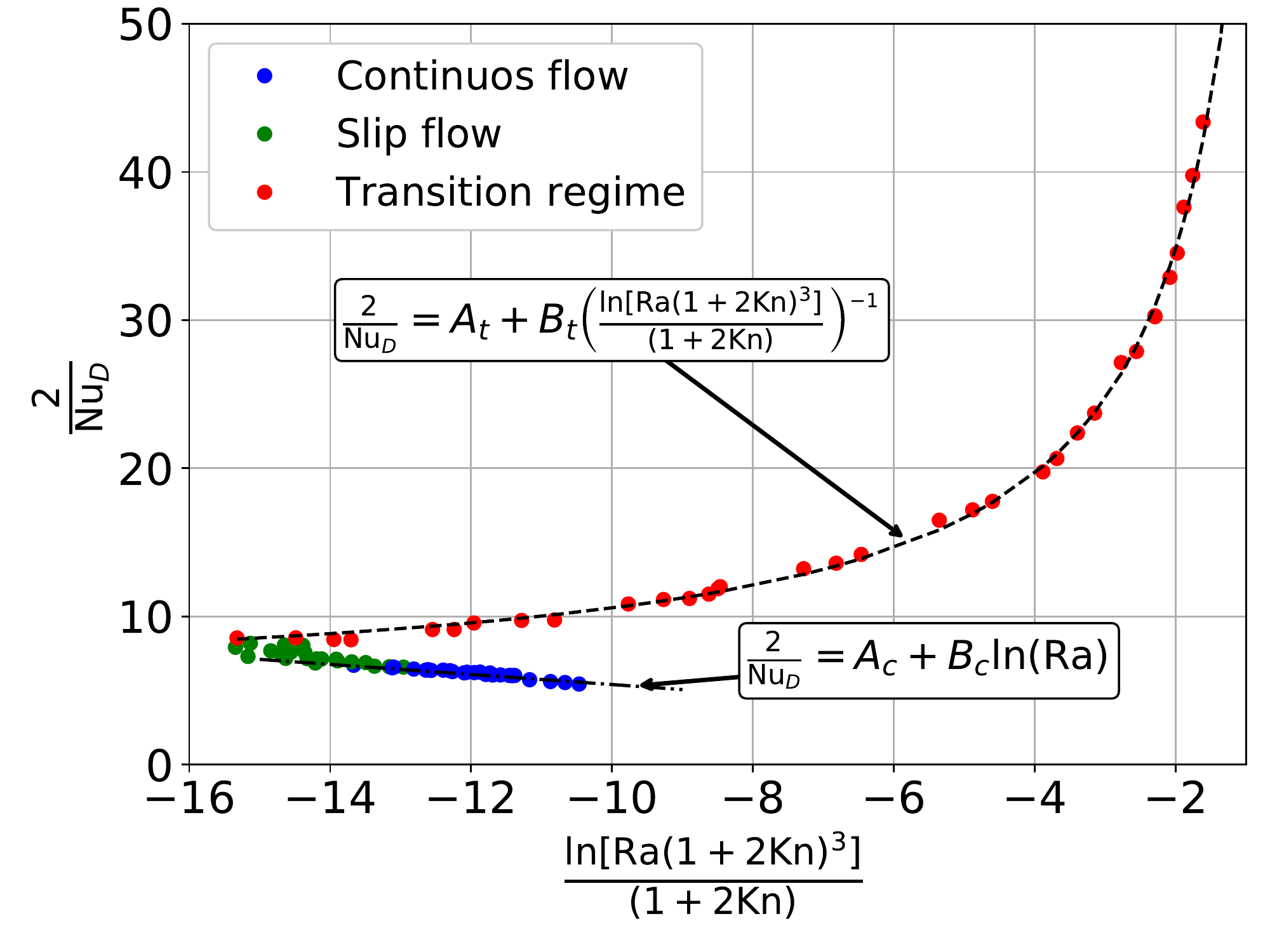}
	\caption{Experimental results of $2/\Nu_D$ for the {$25$~$\mu\mathrm{m}$} wire are plotted against a function of $\Ra$ and $\Kn$. The different behaviour between continuum and transition regimes is exposed.}
	\label{fig:corr_trans}
\end{figure}

The experimental values of $2/\Nu_{D}$ are plotted versus {$\ln[\Ra(1+2\Kn)^3]/(1+2\Kn)$} in Figure \ref{fig:corr_trans}, and a completely different trend is observed between the continuum and the transition regimes. In the continuum regime $\Kn$ is very small, thus
$$\dfrac{\ln[\Ra(1+2\Kn)^3]}{(1+2\Kn)} \rightarrow \ln(\Ra),$$
and the correlation for the continuum regime from equation~(\ref{eq:corr_cont}) is recovered. The presented experimental points for continuum regime collapse on a straight line as explained above. The experimental points for transition regime have the opposite trend, as their values are increasing for increasing values of horizontal axis. The curve formed by the points seems to approach a rectangular hyperbola. Thus, the correlation proposed for the transition regime is of the form
\begin{equation}
\frac{2}{\Nu_{D,t}} = A_t + B_t\dfrac{(1+2\Kn)}{\ln[\Ra(1+2\Kn)^3]},
\label{eq:corr_trans}
\end{equation}

\begin{figure}[h]
	\includegraphics[width=\mywidth]{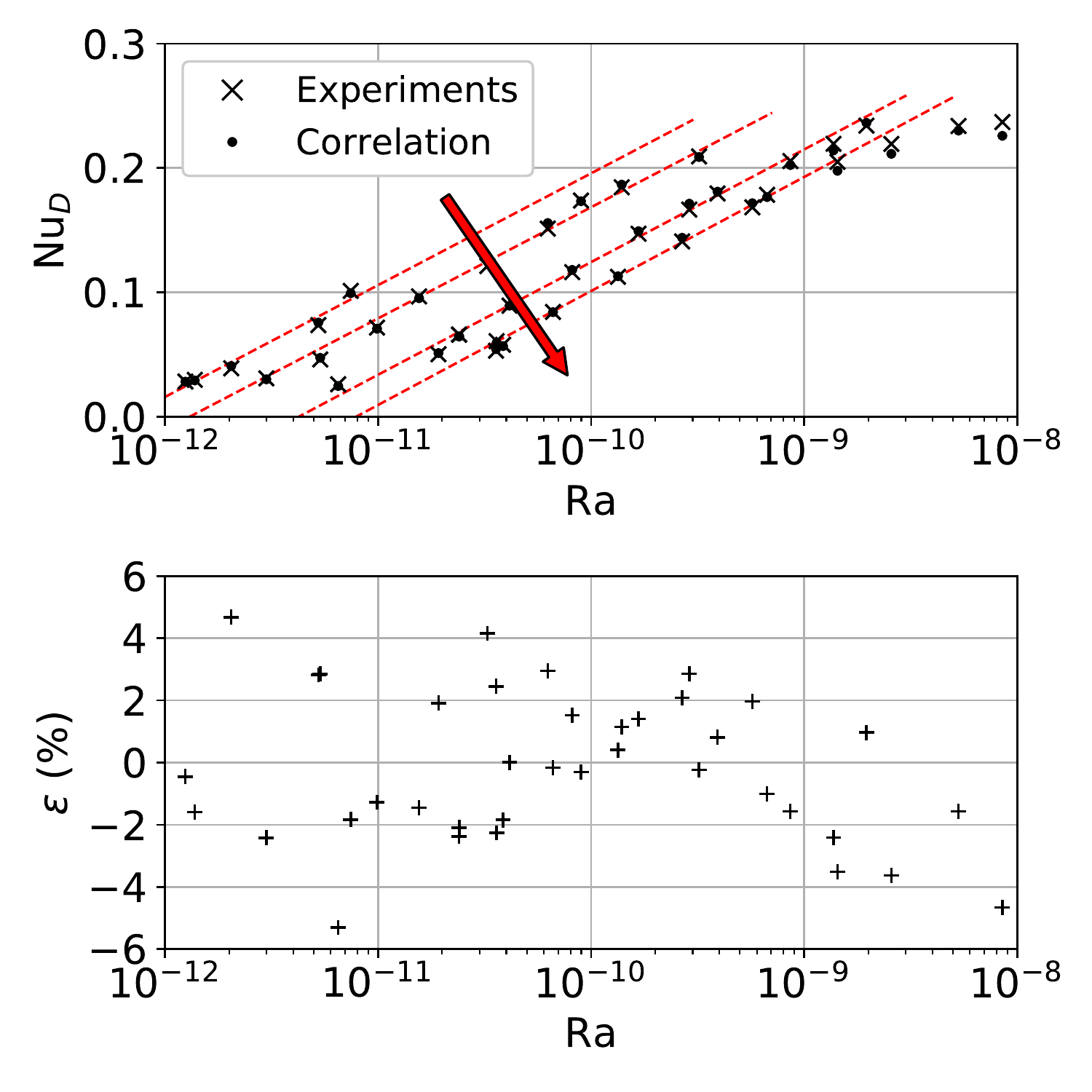}
	\centering
	\caption{(\textit{Top})Experimental and calculated values of $\Nu_D$ as a function of $\Ra$ in the range {$10^{-12}<\Ra<10^{-8}$} for the {$25$~$\mu\mathrm{m}$} diameter wire. \textit{Red dashed lines}  represent approximately constant values of supplied electrical power to the wire: $1.8,\ 3.8,\ 11.1$ and {$19.6$~$\mathrm{mW}$}. Increasing values indicated by red arrow.(\textit{Bottom})The relative error between the experimental and calculated values presented above calculated as indicated in equation~(\ref{eq:error}).}
	\label{fig:error}
\end{figure}

The constants for correlation presented in equation~(\ref{eq:corr_trans}) are also obtained from adjusting the experimental data by least mean squares, and their values are $A_t = 4.5$ and $B_t = -60.7$. The proposed correlation is valid only in the range $0.1<\Kn<10$. In {Figure~\ref{fig:error}} the experimental data are presented together with the values of the Nusselt number calculated using the correlation of equation~(\ref{eq:corr_trans}) and it can be observed that the relative error, defined as
\begin{equation}
\varepsilon = \dfrac{\Nu_{D,exp} - \Nu_{D,corr}}{\Nu_{D,corr}},
\label{eq:error}
\end{equation}
is small, indeed for all experimental points this error is below $6\%$.

Using both proposed correlations, the Nusselt number for very thin wires can be calculated up to a maximum Knudsen number equal to $10$. However, the switch between the correlations is not well-defined, and depending on the problem a sharp change in the Nusselt number can show up if the only changing criterion between the two regimes is the value of the Knudsen number. For the purpose of achieving a better approximation, it is proposed the utilization of the logistic function with the form
\begin{equation}
\mathcal{L} = \dfrac{1}{1+\exp\left(-\dfrac{\Kn - 0.2}{0.01}\right)}.
\end{equation}
Therefore the complete correlation using the logistic function is
\begin{equation}
\dfrac{2}{\Nu_D} = \mathcal{L}\left( 4.5 - 60.7 \dfrac{(1+2\Kn)}{\ln[\Ra(1+2\Kn)^3]}\right) + (1-\mathcal{L})[2 - 0.34\ln(\Ra)].
\label{eq:corr}
\end{equation}
Its validity range is $10^{-12}< \Ra < 1$ and $\Kn < 10$.

With regard to the validity range of the Rayleigh number, in principle, based on our data, the validity of the correlation would be just for Rayleigh numbers below $10^{-4}$. However, it has been verified that the Nusselt number calculated using {equation~(\ref{eq:corr_cont})} is between the values obtained using correlations~\cite{kyte1953} and~\cite{fujii1982} up to Rayleigh numbers above unity. This allows for the expansion of the validity range of the {equation~(\ref{eq:corr})} up to that point $\Ra<1$. In addition, the validity range can be extensively increased by replacing our correlation for the continuum regime with the correlation proposed by \citet{fujii1982}. In this case, as stated by the cited authors, the validity range would be $10^{-12}< \Ra < 10^{6}$, although the Knudsen number would remain below $10$.

\begin{figure}[h]
	\includegraphics[width=\mywidth]{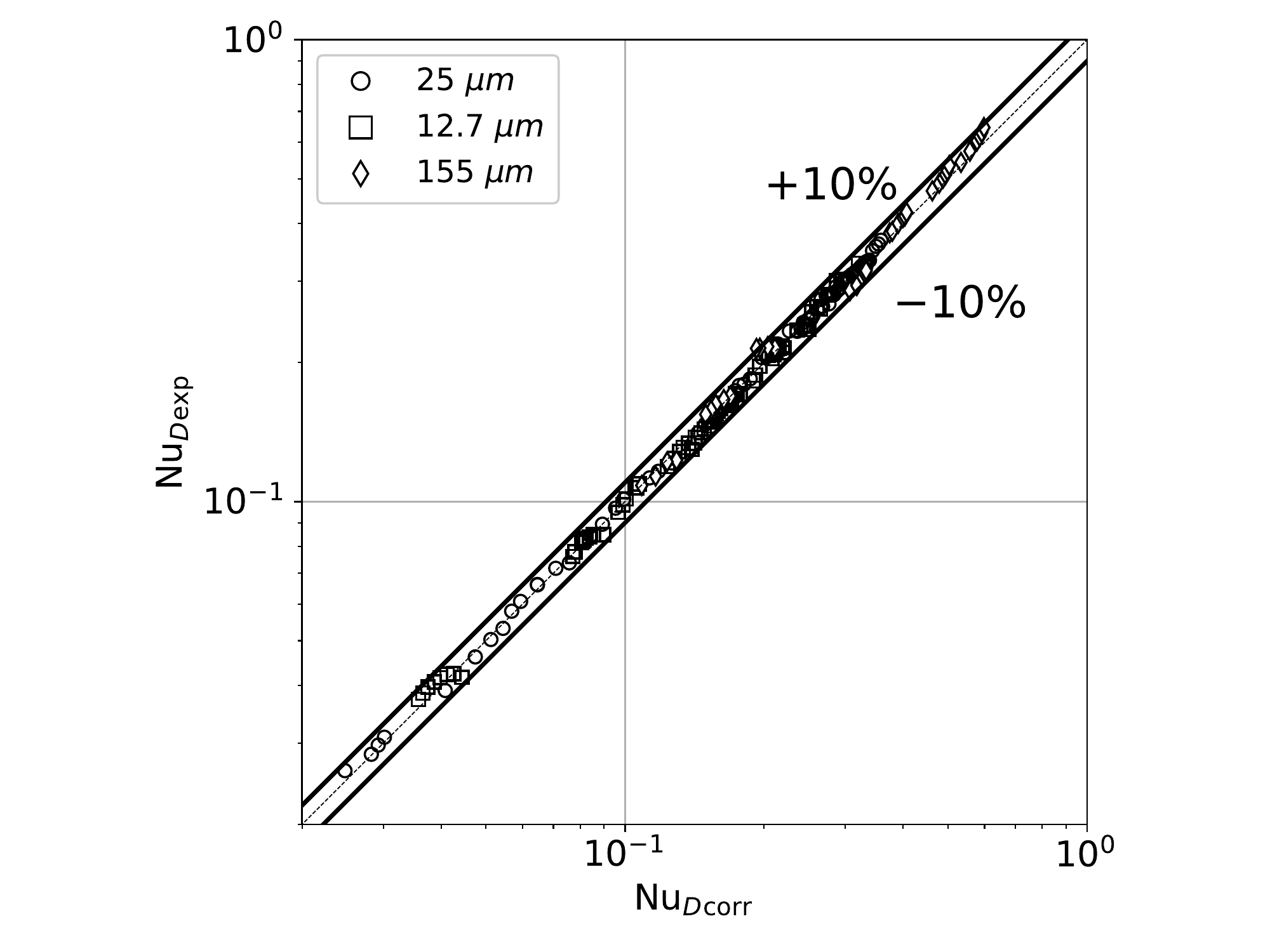}
	\centering
	\caption{Comparison between experimental data and calculated values of $\Nu_D$ using the correlation proposed in equation~(\ref{eq:corr}). For a given pair of experimental values of $\Ra$ and $\Kn$, a ${\Nu_D}_{\mathrm{exp}}$ is known (vertical axis) and a ${\Nu_D}_{\mathrm{corr}}$ is calculated (horizontal axis). These two values are the coordinates of the point in the presented graph. The diagonal of the graph (dashed line) corresponds to a perfect match between experiments and correlation, and the two solid lines mark the relative error of $\pm 10\%$ between them. Data for the $155\ \mu\mathrm{m}$ wire are extracted from~\cite{milano2010}.}
	\label{fig:valid}
\end{figure}

The presented correlation was obtained from experimental data of the $25\ \mu\mathrm{m}$ wire. The lack of experimental data for very fine wires in rarefied atmospheres, makes it difficult to validate of the correlations by comparing with other authors. For the purpose of its validation, several experiments with the \mbox{$12.7\ \mu\mathrm{m}$} wire were carried out. The final results were obtained using the same process explained in section~\ref{sec:prob}. Finally, the Nusselt numbers obtained experimentally are plotted against the ones predicted by the proposed correlation in \mbox{Figure~\ref{fig:valid}}. In this graph, for a given pair of experimental values of $\Ra$ and $\Kn$, a ${\Nu_D}_{\mathrm{exp}}$ is known and a ${\Nu_D}_{\mathrm{corr}}$ is calculated using correlation (\ref{eq:corr}). These two values are the coordinates of the point in the presented graph. The diagonal of the graph corresponds to a perfect match between experiments and correlation, and the parallel lines at both sides of the diagonal mark the relative error of $\pm 10\%$ between them. Experimental results for a copper wire with a \mbox{$155\ \mu \mathrm{m}$} diameter extracted from~\cite{milano2010}, which contains data for the transition regime, are included. The maximum relative error for all the points is less than $10\%$, which denotes that the correlation presented in  equation~(\ref{eq:corr}) is in good agreement with the experimental data.

\section{Conclusions}
An experimental study of free convection in rarefied gases over very thin horizontal wires is presented. The experiments were conducted in a vacuum chamber filled with air between pressures from \mbox{$0.03$~mbar} to ambient pressure, using alumel specimen wires with diameters of \mbox{$12.7\ \mu\mathrm{m}$} and \mbox{$25\ \mu\mathrm{m}$}, and aspect ratio $L/D$ of $5040$ and $2440$ respectively. The estimation of free convection on the wire accounts for thermal radiation and heat conduction losses  to the wire's supports. The obtained values of dimensionless parameters, $\Nu_D$ and $\Ra$, are in accordance with the correlations published in the literature for slip flow and continuum regimes. In transition regime, a noticeable change in the trend of $\Nu_D$ as a function of $\Ra$ is evidenced  in comparison with the cases of slip flow and continuum regimes, characterised by a remarkable reduction in Nusselt number.  Moreover, a particular behaviour observed in transition regime indicates the need to include $\Kn$ in the correlation apart from the usual dependence of $\Ra$. Using the data from the \mbox{$25\ \mu\mathrm{m}$} wire a new correlation, which covers the range from transition to continuum regimes for $\Kn < 10$ and Rayleigh values  $10^{-12}< \Ra < 1$ is proposed. The correlation is validated using data from the \mbox{$12.7\ \mu\mathrm{m}$} wire and from a \mbox{$155\ \mu\mathrm{m}$} wire published by \citet{milano2010}, which include experimental data for transition regime. This correlation could be applicable for very thin horizontal wires (with diameters in the order of micrometers)  in rarefied gases such as Mars surface atmosphere or other low pressure problems.

\section*{Declaration of interests}
The authors declare that they have no known competing financial interests or personal relationships that could have appeared to influence the work reported in this paper.

\section*{Acknowledgements}
Financial support for this work was provided by the Spanish Ministerio de Ciencia e Innovación, Projects {ESP2016-77548-C5-3-R} and {RTI2018-096886-B-C55}.

The authors wish to thank Professor Ángel Sanz Andrés for his constant support on the work progress and constructive comments for the manuscript, Professor Isidoro Martínez Herranz for his support on characterizing materials and Alejandro Fernández Herrero for programming the measurement software. Also, we thank IDR staff for their support on the laboratory instrumentation and their appreciated advices.
\section*{Supplementary material}
Supplementary data associated with this article can be found at \href{http://dx.doi.org/10.17632/mvmcpwvs4k.1}{http://dx.doi.org/\- 10.17632/\-mvmcpwvs4k.1}.

\bibliography{mybibfile}

\begin{thebibliography}{25}
\expandafter\ifx\csname natexlab\endcsname\relax\def\natexlab#1{#1}\fi
\providecommand{\url}[1]{\texttt{#1}}
\providecommand{\href}[2]{#2}
\providecommand{\path}[1]{#1}
\providecommand{\DOIprefix}{doi:}
\providecommand{\ArXivprefix}{arXiv:}
\providecommand{\URLprefix}{URL: }
\providecommand{\Pubmedprefix}{pmid:}
\providecommand{\doi}[1]{\href{http://dx.doi.org/#1}{\path{#1}}}
\providecommand{\Pubmed}[1]{\href{pmid:#1}{\path{#1}}}
\providecommand{\bibinfo}[2]{#2}
\ifx\xfnm\relax \def\xfnm[#1]{\unskip,\space#1}\fi
\bibitem[{G{\'o}mez-Elvira et~al.(2012)G{\'o}mez-Elvira, Armiens, Casta{\~n}er,
  Dom{\'\i}nguez, Genzer, G{\'o}mez, Haberle, Harri, Jim{\'e}nez,
  Kahanp{\"a}{\"a} et~al.}]{gomez2012}
\bibinfo{author}{J.~G{\'o}mez-Elvira}, \bibinfo{author}{C.~Armiens},
  \bibinfo{author}{L.~Casta{\~n}er}, \bibinfo{author}{M.~Dom{\'\i}nguez},
  \bibinfo{author}{M.~Genzer}, \bibinfo{author}{F.~G{\'o}mez},
  \bibinfo{author}{R.~Haberle}, \bibinfo{author}{A.-M. Harri},
  \bibinfo{author}{V.~Jim{\'e}nez}, \bibinfo{author}{H.~Kahanp{\"a}{\"a}},
  et~al.,
\newblock \bibinfo{title}{Rems: The environmental sensor suite for the mars
  science laboratory rover},
\newblock \bibinfo{journal}{Space science reviews} \bibinfo{volume}{170}
  (\bibinfo{year}{2012}) \bibinfo{pages}{583--640}.
  \DOIprefix\doi{10.1007/s11214-012-9921-1}.
\bibitem[{P{\'e}rez-Grande et~al.(2011)P{\'e}rez-Grande, Sanz-Andr{\'e}s,
  Bezdenejnykh, Farrahi, Barthol, and Meller}]{perez2011}
\bibinfo{author}{I.~P{\'e}rez-Grande}, \bibinfo{author}{A.~Sanz-Andr{\'e}s},
  \bibinfo{author}{N.~Bezdenejnykh}, \bibinfo{author}{A.~Farrahi},
  \bibinfo{author}{P.~Barthol}, \bibinfo{author}{R.~Meller},
\newblock \bibinfo{title}{Thermal control of sunrise, a balloon-borne solar
  telescope},
\newblock \bibinfo{journal}{Proceedings of the Institution of Mechanical
  Engineers, Part G: Journal of Aerospace Engineering} \bibinfo{volume}{225}
  (\bibinfo{year}{2011}) \bibinfo{pages}{1037--1049}.
  \DOIprefix\doi{10.1177/0954410011401711}.
\bibitem[{Gao et~al.(2018)Gao, Xie, Xiong, and Yue}]{gao2018}
\bibinfo{author}{J.~Gao}, \bibinfo{author}{D.~Xie}, \bibinfo{author}{Y.~Xiong},
  \bibinfo{author}{Y.~Yue},
\newblock \bibinfo{title}{Thermal characterization of microscale heat
  convection in rare-gas environment by a steady-state “hot wire” method},
\newblock \bibinfo{journal}{Applied Physics Express} \bibinfo{volume}{11}
  (\bibinfo{year}{2018}) \bibinfo{pages}{066601}.
  \DOIprefix\doi{10.7567/APEX.11.066601}.
\bibitem[{Tsien(1946)}]{tsien1946}
\bibinfo{author}{H.-S. Tsien},
\newblock \bibinfo{title}{Superaerodynamics, mechanics of rarefied gases},
\newblock \bibinfo{journal}{Journal of the Aeronautical Sciences}
  \bibinfo{volume}{13} (\bibinfo{year}{1946}) \bibinfo{pages}{653--664}.
  \DOIprefix\doi{10.2514/8.11476}.
\bibitem[{Springer(1971)}]{springer1971}
\bibinfo{author}{G.~S. Springer},
\newblock \bibinfo{title}{Heat transfer in rarefied gases},
\newblock in: \bibinfo{booktitle}{Advances in heat transfer},
  volume~\bibinfo{volume}{7}, \bibinfo{publisher}{Elsevier},
  \bibinfo{year}{1971}, pp. \bibinfo{pages}{163--218}.
  \DOIprefix\doi{10.1016/S0065-2717(08)70018-2}.
\bibitem[{Davis and Davies(1972)}]{davis1972}
\bibinfo{author}{M.~Davis}, \bibinfo{author}{P.~Davies},
\newblock \bibinfo{title}{Factors influencing the heat transfer from
  cylindrical anemometer probes},
\newblock \bibinfo{journal}{International Journal of Heat and Mass Transfer}
  \bibinfo{volume}{15} (\bibinfo{year}{1972}) \bibinfo{pages}{1659--1677}.
  \DOIprefix\doi{10.1016/0017-9310(72)90096-8}.
\bibitem[{Perez-Grande et~al.(2017)Perez-Grande, Peinado, Chamorro, Torralbo,
  Alonso, Rodriguez~Manfredi, Lepinette, and Sebastian}]{perez2017}
\bibinfo{author}{I.~Perez-Grande}, \bibinfo{author}{L.~Peinado},
  \bibinfo{author}{A.~Chamorro}, \bibinfo{author}{I.~Torralbo},
  \bibinfo{author}{G.~Alonso}, \bibinfo{author}{J.~A. Rodriguez~Manfredi},
  \bibinfo{author}{A.~Lepinette}, \bibinfo{author}{E.~Sebastian},
\newblock \bibinfo{title}{Thermal design of the {A}ir {T}emperature {S}ensor
  ({ATS}) and the {T}hermal {I}nfra{R}ed {S}ensor ({TIRS}) of the {M}ars
  {E}nvironmental {D}ynamics {A}nalyzer ({MEDA}) for {M}ars 2020},
\newblock \bibinfo{organization}{47th International Conference on Environmental
  Systems}, \bibinfo{year}{2017}.
\bibitem[{Kyte et~al.(1953)Kyte, Madden, and Piret}]{kyte1953}
\bibinfo{author}{J.~Kyte}, \bibinfo{author}{A.~Madden}, \bibinfo{author}{E.~L.
  Piret},
\newblock \bibinfo{title}{Natural-convection heat transfer at reduced
  pressure-spheres and cylinders},
\newblock \bibinfo{journal}{Chemical Engineering Progress} \bibinfo{volume}{49}
  (\bibinfo{year}{1953}) \bibinfo{pages}{653--662}.
\bibitem[{Collis and Williams(1954)}]{collis1954}
\bibinfo{author}{D.~Collis}, \bibinfo{author}{M.~Williams},
  \bibinfo{title}{Free convection of heat from fine wires},
  \bibinfo{publisher}{Department of Supply, Aeronautical Research
  Laboratories}, \bibinfo{year}{1954}.
\bibitem[{Mahony(1957)}]{mahony1957}
\bibinfo{author}{J.~Mahony},
\newblock \bibinfo{title}{Heat transfer at small grashof numbers},
\newblock \bibinfo{journal}{Proceedings of the Royal Society of London. Series
  A. Mathematical and Physical Sciences} \bibinfo{volume}{238}
  (\bibinfo{year}{1957}) \bibinfo{pages}{412--423}.
  \DOIprefix\doi{10.1098/rspa.1957.0009}.
\bibitem[{Fujii et~al.(1979)Fujii, Fujii, and Matsunaga}]{fujii1979}
\bibinfo{author}{T.~Fujii}, \bibinfo{author}{M.~Fujii},
  \bibinfo{author}{T.~Matsunaga},
\newblock \bibinfo{title}{A numerical analysis of laminar free convection
  around an isothermal horizontal circular cylinder},
\newblock \bibinfo{journal}{Numerical Heat Transfer, Part A: Applications}
  \bibinfo{volume}{2} (\bibinfo{year}{1979}) \bibinfo{pages}{329--344}.
  \DOIprefix\doi{10.1080/10407787908913417}.
\bibitem[{Fujii et~al.(1982)Fujii, Fujii, and Honda}]{fujii1982}
\bibinfo{author}{M.~Fujii}, \bibinfo{author}{T.~Fujii},
  \bibinfo{author}{T.~Honda},
\newblock \bibinfo{title}{Theoretical and experimental studies of the free
  convection around a long horizontal thin wire in air},
\newblock in: \bibinfo{booktitle}{International Heat Transfer Conference
  Digital Library}, \bibinfo{organization}{Begel House Inc.},
  \bibinfo{year}{1982}, pp. \bibinfo{pages}{311--316}.
  \DOIprefix\doi{10.1615/IHTC7.3300}.
\bibitem[{Xie et~al.(2017)Xie, Li, Liu, Wang, and Wang}]{xie2017}
\bibinfo{author}{F.~Xie}, \bibinfo{author}{Y.~Li}, \bibinfo{author}{Z.~Liu},
  \bibinfo{author}{X.~Wang}, \bibinfo{author}{L.~Wang},
\newblock \bibinfo{title}{A forced convection heat transfer correlation of
  rarefied gases cross-flowing over a circular cylinder},
\newblock \bibinfo{journal}{Experimental Thermal and Fluid Science}
  \bibinfo{volume}{80} (\bibinfo{year}{2017}) \bibinfo{pages}{327--336}.
  \DOIprefix\doi{10.1016/j.expthermflusci.2016.09.002}.
\bibitem[{Xie et~al.(2018)Xie, Li, Wang, Wang, Lei, and Xing}]{xie2018}
\bibinfo{author}{F.~Xie}, \bibinfo{author}{Y.~Li}, \bibinfo{author}{X.~Wang},
  \bibinfo{author}{Y.~Wang}, \bibinfo{author}{G.~Lei},
  \bibinfo{author}{K.~Xing},
\newblock \bibinfo{title}{Numerical study on flow and heat transfer
  characteristics of low pressure gas in slip flow regime},
\newblock \bibinfo{journal}{International Journal of Thermal Sciences}
  \bibinfo{volume}{124} (\bibinfo{year}{2018}) \bibinfo{pages}{131--145}.
  \DOIprefix\doi{10.1016/j.ijthermalsci.2017.09.022}.
\bibitem[{Morgan(1975)}]{morgan1975}
\bibinfo{author}{V.~T. Morgan},
\newblock \bibinfo{title}{The overall convective heat transfer from smooth
  circular cylinders},
\newblock in: \bibinfo{booktitle}{Advances in heat transfer},
  volume~\bibinfo{volume}{11}, \bibinfo{publisher}{Elsevier},
  \bibinfo{year}{1975}, pp. \bibinfo{pages}{199--264}.
  \DOIprefix\doi{10.1016/S0065-2717(08)70075-3}.
\bibitem[{Churchill and Chu(1975)}]{churchill1975}
\bibinfo{author}{S.~W. Churchill}, \bibinfo{author}{H.~H. Chu},
\newblock \bibinfo{title}{Correlating equations for laminar and turbulent free
  convection from a horizontal cylinder},
\newblock \bibinfo{journal}{International journal of heat and mass transfer}
  \bibinfo{volume}{18} (\bibinfo{year}{1975}) \bibinfo{pages}{1049--1053}.
  \DOIprefix\doi{10.1016/0017-9310(75)90222-7}.
\bibitem[{Boetcher(2014)}]{boetcher2014}
\bibinfo{author}{S.~K. Boetcher}, \bibinfo{title}{Natural convection from
  circular cylinders}, \bibinfo{publisher}{Springer}, \bibinfo{year}{2014}.
  \DOIprefix\doi{10.1007/978-3-319-08132-8_2}.
\bibitem[{Seo et~al.(2019)Seo, Park, Choi, Jo, and Yoon}]{seo2019}
\bibinfo{author}{M.-H. Seo}, \bibinfo{author}{J.-H. Park},
  \bibinfo{author}{K.-W. Choi}, \bibinfo{author}{M.-S. Jo},
  \bibinfo{author}{J.-B. Yoon},
\newblock \bibinfo{title}{An investigation of surficial conduction heat loss in
  perfectly aligned micro-wire array},
\newblock \bibinfo{journal}{Applied Physics Letters} \bibinfo{volume}{115}
  (\bibinfo{year}{2019}) \bibinfo{pages}{131901}.
  \DOIprefix\doi{10.1063/1.5123523}.
\bibitem[{Kasap et~al.(2017)Kasap, Koughia, and Ruda}]{kasap2017}
\bibinfo{author}{S.~Kasap}, \bibinfo{author}{C.~Koughia},
  \bibinfo{author}{H.~E. Ruda},
\newblock \bibinfo{title}{Electrical conduction in metals and semiconductors},
\newblock in: \bibinfo{booktitle}{Springer Handbook of Electronic and Photonic
  Materials}, \bibinfo{publisher}{Springer}, \bibinfo{year}{2017}, pp.
  \bibinfo{pages}{1--1}.
\bibitem[{Incropera et~al.(2007)Incropera, Lavine, Bergman, and
  DeWitt}]{incropera}
\bibinfo{author}{F.~P. Incropera}, \bibinfo{author}{A.~S. Lavine},
  \bibinfo{author}{T.~L. Bergman}, \bibinfo{author}{D.~P. DeWitt},
  \bibinfo{title}{Fundamentals of heat and mass transfer},
  \bibinfo{publisher}{Wiley}, \bibinfo{year}{2007}.
\bibitem[{Collis and Williams(1959)}]{collis1959}
\bibinfo{author}{D.~Collis}, \bibinfo{author}{M.~Williams},
\newblock \bibinfo{title}{Two-dimensional convection from heated wires at low
  reynolds numbers},
\newblock \bibinfo{journal}{Journal of Fluid Mechanics} \bibinfo{volume}{6}
  (\bibinfo{year}{1959}) \bibinfo{pages}{357--384}.
  \DOIprefix\doi{10.1017/S0022112059000696}.
\bibitem[{Milano(2010)}]{milano2010}
\bibinfo{author}{G.~Milano},
\newblock \bibinfo{title}{Some experimental results of free convection from
  horizontal heated wires below atmospheric pressure},
\newblock in: \bibinfo{booktitle}{2010 14th International Heat Transfer
  Conference}, \bibinfo{organization}{American Society of Mechanical Engineers
  Digital Collection}, \bibinfo{year}{2010}, pp. \bibinfo{pages}{183--192}.
\bibitem[{Sasaki et~al.(1994)Sasaki, Masuda, Higano, and
  Hishinuma}]{sasaki1994}
\bibinfo{author}{S.~Sasaki}, \bibinfo{author}{H.~Masuda},
  \bibinfo{author}{M.~Higano}, \bibinfo{author}{N.~Hishinuma},
\newblock \bibinfo{title}{Simultaneous measurements of specific heat and total
  hemispherical emissivity of chromel and alumel by a transient calorimetric
  technique},
\newblock \bibinfo{journal}{International journal of thermophysics}
  \bibinfo{volume}{15} (\bibinfo{year}{1994}) \bibinfo{pages}{547--565}.
\bibitem[{{Engineering ToolBox}(2009)}]{tablas}
\bibinfo{author}{{Engineering ToolBox}}, \bibinfo{title}{{Air - Thermal
  Conductivity}}, \bibinfo{year}{2009}. \URLprefix
  \url{{https://www.engineeringtoolbox.com/air-properties-viscosity-conductivity-heat-capacity-d_1509.html}},
  \bibinfo{note}{{Accessed: 2018-03-01}}.
\bibitem[{Beckwith et~al.(2009)Beckwith, Marangoni, and
  Lienhard}]{beckwith2009}
\bibinfo{author}{T.~G. Beckwith}, \bibinfo{author}{R.~D. Marangoni},
  \bibinfo{author}{J.~H. Lienhard}, \bibinfo{title}{Mechanical measurements},
  \bibinfo{publisher}{Pearson}, \bibinfo{year}{2009}.

\end{thebibliography}

\end{document}